\documentstyle[preprint,aps]{revtex}
\begin{document} 
\title{The Isovector Quadrupole-Quadrupole Interaction Used in Shell
Model Calculations with $\Delta N=2$ Excitations}
 
\author{M.S. Fayache,\cite{byline}
Y.Y. Sharon and L. Zamick}
\address{Department of Physics and Astronomy, Rutgers University,
Piscataway, New Jersey 08855} 
\date{\today}
\maketitle

\begin{abstract}
The effect of the ``$B$'' term in the interaction $-\chi Q\cdot
Q(1+B\vec{\tau}(1)\cdot \vec{\tau}(2))$, which was previously
considered in the $0p$ shell, is now studied in a larger space which
includes $\Delta N=2$ excitations. We still get a collapse of low-lying
states below the conventional $J=0^+$ ground state when $B$ is made
sufficiently negative. 
\end{abstract}
 
\narrowtext

\section{Introduction}

In a previous work~\cite{qqt}, a combination of the isoscalar and the
isovector quadrupole-quadrupole interaction $-\chi Q\cdot
Q(1+B\vec{\tau}(1)\cdot \vec{\tau}(2))$ was used in a shell model
calculation of states in $^{10}Be$. The model space was the $0p$
shell, and in particular the dependence on the parameter $B$ (the
strength of the isovector term in the interaction) was studied. In the
present work, we extend the calculations to a space which includes the
$0p$ shell plus all $\Delta N=2$ excitations. 

For $B=0$ we have a spin-isospin independent interaction, and for the
$0p$ shell the Wigner Supermultiplet Theory applies~\cite{wig}. We can
also think of this as an application of the Elliott model
\cite{Elliott} to the $0p$ shell. Of course, the greatest interest in
the latter model will be in the $1s-0d$ shell. The study of the
addition of the term $-\chi B\vec{\tau}(1)\cdot \vec{\tau}(2)$ can be
regarded as an exploration of what happens when we deviate from the
simple $SU(3)$ limit. In another vein, a large negative value of $B$
has been invoked in $R.P.A.$ calculations to explain the splitting of
the isovector and isoscalar giant quadrupole resonances. Will this
same large negative $B$ give better results than the $B=0$, $SU(3)$
limit in shell model calculations?  

In the previous calculation of $^{10}Be$ in the $0p$ shell~\cite{qqt},
the answer was a resounding no! In the $0p$ space calculation with
$B=0$, the ground state has $J=0^+$ and orbital symmetry [42]. There
is a two-fold degeneracy for the first excited state: two $J=2^+$
states both with orbital symmetry [42], corresponding to $K=0$ and
$K=2$. The focus of interest was also on two degenerate $L=1~S=1$
states: one with orbital symmetry [33], and the other [411]. From
$L=1~S=1$ one can get a triplet of states $J=0,~1,~2$. 

When a negative $B$ of increasing magnitude was introduced, the $2_1$
state came down in energy in a nearly linear fashion. The $B(E2)$ to
this state was purely isovector i.e. proportional to ($e_p -
e_n$)$^2$. Ultimately, at $B \approx -0.68$ this $2^+$ state became
the ground state.  

However, coming down even faster was a triplet of states with
$J=0^+,~1^+$ and $2^+$ which for $B=0$ coincided with the degenerate
[33] and [411] $L=1~S=1$ states mentioned above. Very quickly, after
the $2^+_1$ state became the ground state, this triplet (which is some
linear combination of states of symmetry [33] and [411]) took over and
became the ground state for $B \leq -0.74$. 

In the following sections we present new calculated results for
$^6He$, $^8Be$, $^{10}Be$ and $^{20}Ne$. For $^8Be$ and $^{10}Be$ the
model spaces are extended to include $2 \hbar \omega$ excitations. The
values of $\chi$ (in $MeV fm^4$) used for the four nuclei above are
respectively 0.6914, 0.57619, 0.36146 and 0.13403.
Another convenient parameter is $\bar{\chi}=\chi \frac{5b^4}{32\pi}$ 
($b^2=\frac{\hbar}{m\omega}$). The corresponding values of $\bar{\chi}$
are 0.200, 0.1865, 0.1286, 0.0691. 

\section{Present Calculations- $^{10}Be$}

In this work, we extend the previous $0p$ space calculation of
$^{10}Be$ to include all $\Delta N=2$ excitations. We can quickly
assert that, although there are some differences, in the most
important respect the new results are remarkably similar to the $0p$
results albeit on a larger scale of $B$. 

For $B=0$ (i.e. a simple $Q \cdot Q$ interaction), it is interesting
to note that when we extend the calculation to $\Delta N=2$, some
degeneracies are preserved and some are not. The supermultiplet
degeneracies are preserved, i.e. for a given [f] the various ($S,T$)
combinations are still degenerate. However, the `accidental' 
degeneracies in the $0p$ space are not maintained in the larger
space. 

For example, the $2_1^+$ and $2^+_2$ states, which were degenerate in
the $0p$ space, are no longer so in the extended space. The $2_1^+$
state is calculated to be at 2.186 $MeV$, and the $2_2^+$ state at
3.400 $MeV$. Most of the calculated $E2$ strength goes to the $2_2^+$
state.  

The two sets of triplets ($L=1~S=1~J=0^+,~1^+,~2^+$) with orbital
symmetry [33] and [411] were accidentally degenerate in the $0p$
space. In the extended space, the states split in such a way that the
excitation energy of one is about twice that of the other. The reason
for the removal of the accidental degeneracy may be that $\Delta N =2$
mixing is equivalent to having different oscillator frequencies in the
$x,y$ and $z$ directions, as in the Nilsson model. 

We now consider for $^{10}Be$ what happens in the large space as a
function of $B$ i.e. we include not only all configurations in the
$0p$ shell but also all 2 $\hbar \omega$ excitations as well. The
results are shown in Fig.~ 1. In the lower part of the figure we
follow the behaviour of the $2^+_1$ state (dashed curve) and the
$L=1~S=1$ triplet $J=0^+,1^+$ and $2^+$ (solid curve). We find that
although the overall scale has changed the results are quite similar
to those in the small space. That is to say, the $2_1$ state and
$L=1~S=1$ triplet both come down in energy as $B$ becomes more
negative. The $L=1~S=1$ triplet however beats out the $2_1^+$ state
and becomes the ground state at $B \approx -1.6$. In the small space
the $2_1^+$ state first became the ground state at $B=-0.68$ and the
$L=1~S=1$ triplet overtook this state and became the ground state at
$B=-0.74$. But this is a minor difference.  

The main point is the change of scale by somewhat more than a factor
of two. There has been considerable discussion of this in the
literature, e.g. by Bes, Broglia and Nilsson~\cite{bbn}. They point
out that in a small space, the renormalized $B$ is considerably
smaller than the bare $B$. For example, with a bare value $B=-3.6$,
the renormalized value to be used in a small space is only $B \approx
-0.6$. However, in the large space we get the undesirable collapse of
states at less than half the `empirical' value of -3.6. 

Another change in going from the small space to the large space
concerns the isoscalar and isovector $B(E2)$'s. In the small space the
$2_1^+$, which came down in energy as $B$ was made increasingly
negative, was a purely isovector state. That is to say the value of
$B(E2,e_p,e_n)$ was zero if $e_p=e_n$ but was large if $e_p=-e_n$. In
the large space the corresponding $2^+$ state is not purely
isovector. At $B=0$, the $2^+_1$ state is at 2.186 $MeV$ and the
$2_2^+$ state at 3.400 $MeV$. The calculated isoscalar $B(E2)$'s
$(e_p=e_n=1)$ are respectively 63.76 and 113.34 $e^2fm^4$; the
isovector $B(E2)$'s $(e_p=1, e_n=-1)$ are respectively 12.49 and 9.52
$e^2fm^4$.  

When $B=-1.5$ (just before the collapse), the corresponding states are
at 0.717 and 4.199 $MeV$. The isoscalar $B(E2)$'s are 32.86 and 168.4
$e^2fm^4$ respectively; the isovector $B(E2)$'s are 13.28 and 0.193
$e^2fm^4$. (At $B=-1.5$ these states are the second and third $2^+$
states. The lowest $2^+$ state is at 0.370 $MeV$ and is part of the
$L=1~S=1$ triplet -it has zero isoscalar and zero isovector strength). 

In the upper part of Fig.~ 1, we also look at the $B$ dependence of
the $T=1$ and $T=2$ branches of the scissors mode (dot-dashed curve
and dotted curve respectively). We note that the energies of both rise
rapidly with increasing negative $B$. This behaviour is somewhat
different from that in the small space, where the $T=2$ branch came
down in energy and the $T=1$ branch came up in energy as $B$ was made
more negative (Fig.~ 1 in Ref.~\cite{qqt}).

It should be pointed out that in a previous publication~\cite{ann} we
estimated that the $T=2$ scissors mode should be at about 22 $MeV$ in
excitation, so one may argue that if we focus just on this point a 
large negative $B$ is a good thing. However, we see in the lower part
of Fig.~1 the unphysical behaviour of collapse which we discussed
previously. 

\section{Non-Collapse in $^8Be$}

In contrast to the behaviour in $^{10}Be$, we find that no states come
down in energy relative to the $(L=0~S=0)^{J=0^+}$ ground state when
the magnitude of negative $B$ is increased. The results for $^8Be$ in
a large space are given in Fig.~ 2. 

Both the energies of the $2_1^+~T=0$ and $1^+~T=1$ states {\em
increase } as $B$ is made more negative. 

This tells us that it is not enough to look at one nucleus to test the
consequences of a certain interaction. By looking at $^8Be$, we find
no problems with collapsing states, but they are certainly there in
$^{10}Be$ and, as we shall soon see, in $^6Be$. 

\section{The $B$ dependence for the two-particle problem $^6He$
($^6Be$)}

We can gain some insight into why we get states coming down in energy
for negative $B$ by considering the simplest problem of two identical
particles in the $p$ shell $i.e.$ $^6He$ or $^6Be$. This analysis also
applies to two holes $e.g.$ $^{12}Be$.

With a spin-isospin independent interaction, the wavefunctions can be
classified by $LS$ coupling:

\noindent Ground State ($L=0~S=0$)$^{J=0^+}$\\
First Excited State ($L=2~S=0$)$^{J=2^+}$\\
Triplet ($L=1~S=1$)$^{J=0+,1^+,2^+}$\\

\noindent The above states all have isospin $T=1$.

With a $Q \cdot Q$ interaction, the energies are given by
\cite{Elliott} 

\[\langle-\chi Q\cdot Q\rangle_{\lambda~\mu~L}=
\bar{\chi}\left[-4(\lambda^2+\mu^2+\lambda\mu+ 3
(\lambda+\mu))+3L(L+1)\right]\] 

\noindent where $\bar{\chi}=\chi \frac{5b^4}{32\pi}$ with $b$ the
harmonic oscillator length parameter ($b^2=\frac{\hbar}{m\omega}$). 

For the even $L$ states ($\lambda \mu$) is equal to (20); for $L=1$
($\lambda \mu$)=(01). With the interaction $-\chi Q\cdot
Q(1+B\vec{\tau}(1)\cdot \vec{\tau}(2))$, we replace $\bar{\chi}$ by 
$\bar{\chi}(1+B)$ (since $\vec{\tau(1)} \cdot \vec{\tau(2)}=+1$ for
$T=1$). The energies are then 

\noindent $L=0~S=0$: -40$\bar{\chi}(1+B)$\\
$L=2~S=0$: -22$\bar{\chi}(1+B)$\\
$L=1~S=1$: -10$\bar{\chi}(1+B)$\\

\noindent The above expressions are shown in Fig.~ 3. For negative
$B$ with $|B| \leq 1$, the $L=0$ state is the 
lowest in energy, and the $L=1~S=1$ triplet is the highest in
energy. For $B=-1$, all three states are degenerate at zero
energy. For $B$ negative but $|B| > 1$, there is a sign change of
the overall coupling, the $L=1~S=1$ triplet becomes the ground
state and the $L=0~S=1$ state is at the highest energy.

\section{The Energy Weighted Strength as a Function of $B$}

In Tables I, II and III we present the results for the summed
isovector orbital $M1$ strength and the corresponding summed
energy-weighted strength for $^{10}Be$ and $^8Be$. The cases
considered are: $^{10}Be~(T=1 \rightarrow T=1)$, $^{10}Be~(T=1
\rightarrow T=2)$ and $^8Be~(T=1 \rightarrow T=1)$. We give results
for both the low-lying sums and the total sums. The division between
low and high energy is somewhat arbitrary but there seems to be a
sudden sharp rise in the cumulative sums which makes the task easier
than one might at first expect.  

Concerning the energy-weighted strength ($EWS$), we recall that for $B=0$
i.e. for $V=-\chi Q \cdot Q$, there is a sum rule which relates the
isovector orbital $B(M1)$ to the difference of the isoscalar and
isovector $B(E2)$~\cite{zz}:

\[EWS (B=0)= \frac{9}{16\pi}\chi \sum \left
[B(E2,e_p=1,e_n=1)-B(E2,e_p=1,e_n=-1) \right ]\]

We next study the $B$-dependence. We define $R(B)$ as the ratio
$EWS(B)/EWS(0)$. The results in Tables II and III for $R(B)$ can be
fit approximately by the following formulae: 

\[^8Be~~~~ T=0 \rightarrow T=1~~~~R(B)\approx(1-1.7B)\]

\begin{eqnarray*}
^{10}Be~~ & T=1 \rightarrow T=1~~~~R(B)\approx(1-3B)\\
          &~~ T=1 \rightarrow T=2~~~~R(B)\approx(1-1.5B)
\end{eqnarray*}

For negative $B$ the energy-weighted sum rule increases relative to
the case $B=0$, as was noted by Hamamoto and Nazarewicz~\cite{hz}. As
can be seen from Tables II and III, the low $EWS$ do not increase as
rapidly as the total sums, and indeed for the case $^{10}Be~ T=1
\rightarrow T=1$ the low $EWS$ ultimately decreases. 

\section{A Small Space Calculation in $^{22}Ne$}

Just to show that the behaviour in $^{10}Be$ of an isovector $Q \cdot
Q$ interaction is not peculiar to this nucleus alone, we have
performed similar calculations in $^{22}Ne$ -also six valence nucleons
but in the $1s-0d$ shell. In Fig.~ 4 we show the behaviour in small
space ($\Delta N=0$). 

The behaviour in this small space calculation is very similar to the
behaviour in $^{10}Be$. At $B=0$ there are two nearly degenerate $2^+$
states - they would be exactly degenerate if we introduced an
appropriate single-particle splitting between $1s$ and $0d$, but we
make these degenerate in this calculation (that is we don't add the
diagonal term $-\frac{\chi}{2} Q(i) \cdot Q(i)$ to the $Q \cdot Q$
interaction, and we don't include the interaction of $0d$ and $1s$
with the core which will contribute to the single-particle
splitting). 

As $B$ is made negative, one of the $2^+$ states comes
down in energy, but the other $2^+$ state which carries most of the
isoscalar $E2$ strength goes up in energy. At $B=-1$, the $2^+$ state
coming down to 0.323 $MeV$ has an isoscalar $B(E2)$ strength ($e_p=1,
e_n=1$) of 29.20 $e^2 fm^4$ and an almost equal isovector $B(E2)$
strength ($e_p=1, e_n=-1$) of 27.39 $e^2 fm^4$. The $2^+$ state which
goes up in energy ($E=1.737~MeV$ at $B=-1$) has an isoscalar $E2$
strength of 421.50 $e^2 fm^4$ and an isovector $E2$ strength of only
7.17 $e^2 fm^4$. Clearly, the upper $2^+$ state is what we generally
identify as the collective $E2$ state, which with a {\em reasonable
interaction} would be the lowest $2^+$ state. 

But coming down even faster than the $2^+$ state is the $L=1~S=1$
triplet with $J=0^+, 1^+, 2^+$. At $B=-1$ this triplet is at 0.107
$MeV$. The $2^+$ member of this triplet has no isoscalar $E2$ strength
and no isovector $E2$ strength -this is consistent with an $L=0~S=0$
assignment to the ground state and an $L=1~S=1$ assignment to this
excited state. When the magnitude of $B$ is increased slightly ($|B|
\geq 1.005$) this triplet becomes the ground state. This is an
unphysical result and sets a limit on the acceptable magnitude of $B$. 

\section{Closing Remarks}

In this work we set limits on the magnitude of the parameter $B$ in
the interaction $-\chi Q\cdot Q(1+B\vec{\tau}(1)\cdot
\vec{\tau}(2))$. Previous to this work, a popular value of $B$ was
-3.6 -this apparently fit the splitting of the isovector and isoscalar
giant quadrupole resonances in medium-heavy nuclei. We had previously
shown in a small space calculation, and we are now showing in a larger
space calculation which includes $2 \hbar \omega $ excitations, that
such a large value of $B$ leads to a ground state with quite a
different nature from what one generally expects. For example in
$^{10}Be$, instead of an $L=0~S=0$ ground state with $J=0^+$, we find
that for $B < -1.5$ the ground state becomes a triplet
$L=1~S=1~J=0^+,1^+,2^+$. This is clearly not in agreement with
nature. 

It is interesting to note that the states which come down in energy, 
and ultimately come below the $L=0~S=0$ original ground state, are
states which do not have the usual collectivity. The $2^+$ state with
the strongest $B(E2)$ behaves in a reasonable fashion as $B$ is made
increasingly negative. Likewise the scissors mode states
($L=1~S=0~T=1$ and $T=2$) go up in energy. The states which come down
are the $L=1~S=1$ triplet for which there is no $M1$ orbital strength
to the $J=1^+$ member, and no isoscalar or isovector strength to the
$2^+$ member. 

The $2^+$ state which comes down in energy is a bit more complicated
to analyze. Although the overall $B(E2)$ with bare charges
$e_p=1,~e_n=0$ (and one should use bare charges in a calculation which
allows $\Delta N=2$ excitations) is small, there is a cancellation
between a substantial isoscalar and isovector parts. Recall~\cite{qqt}
that in the small space calculation the $2_1^+$ state which comes down
in energy had zero isoscalar $B(E2)$, but the isovector $B(E2)$ (i.e. 
$e_p=1,~e_n=-1$) was very large. 

In Ref.~\cite{qqt} it was pointed out that even in the case of the
isovector-isoscalar splitting of the giant quadrupole resonance, we
could get a smaller value of $B$ (in absolute value) by bringing
effective mass into the picture. Roughly speaking, the unperturbed
energy of the giant resonances, instead of being $2 \hbar \omega$,
will be higher: $2 \hbar \omega/(\frac{m^*}{m})$. Thus, one doesn't need
a $B$ with such a large magnitude in order to get up to the isovector
quadrupole resonance. There have been several works quoted in
Ref.~\cite{qqt} (which will not be repeated here) in which it is
argued that smaller values of $B$ than the reference value of -3.6
would be preferable. Most recently, there has been an analysis by
R. Nojarov of the isovector part of the optical potential, which leads
him to a similar conclusion. 

\section{Acknowledgment}

This work was supported by the Department of Energy Grant No.
DE-FG05-40299. Y.Y. Sharon was on sabbatical leave from J. William
Stockton College. We thank Shelly Sharma for her help and advice and
for her involvement in related works. 

\begin{table}
\caption{The Summed Isovector Orbital Strength in $^{10}Be$ as a
function of $B$ (in units of $\mu_N^2$).}
\begin{tabular}{lrrrr}
  & \multicolumn{2}{c}{$T=1 \rightarrow T=1$} & \multicolumn{2}{c}{$T=1
\rightarrow T=2$}\\
\tableline
$B$ & Low Sum & Total Sum & Low Sum & Total Sum\\
 0   & 0.071   & 0.178   & 0.142   & 0.173\\
-0.5 & 0.085   & 0.311   & 0.149   & 0.207\\
-1.0 & 0.065   & 0.378   & 0.140   & 0.220\\
-1.5 & 0.053   & 0.407   & 0.133   & 0.228\\
\end{tabular}
\end{table}

\begin{table}
\caption{The Energy Weighted Summed Isovector Orbital Strength in
$^{10}Be$ as a function of $B$ (in units of $\mu_N^2$).}
\begin{tabular}{lrrrr}
  & \multicolumn{2}{c}{$T=1 \rightarrow T=1$} & \multicolumn{2}{c}{$T=1
\rightarrow T=2$}\\
\tableline
$B$ & Low Sum & Total Sum & Low Sum & Total Sum\\
  0   & 0.857   & 4.283     & 1.726   & 2.923\\
-0.5  & 1.399   & 10.700    & 2.333   & 5.145\\
-1.0  & 1.198   & 17.213    & 2.609   & 7.273\\
-1.5  & 1.022   & 25.414    & 2.756   & 9.288\\
\end{tabular}
\end{table}

\begin{table}
\caption{The Summed Isovector Orbital Strength (Sum) and Energy-Weighted
Strength (EWS) in $^8Be$ as a function of $B$.}
\begin{tabular}{lrrrr}
  & \multicolumn{2}{c}{$\sum B(M1)~(\mu_N^2)$} & \multicolumn{2}{c}
{$\sum E_xB(M1)~(\mu_N^2 MeV)$}\\
\tableline
$B$ & Low Sum & Total Sum & Low EWS & Total EWS\\
 0   & 0.6225 & 0.7077 & 11.058 & 15.153\\
-0.5 & 0.6033 & 0.8466 & 14.409 & 29.393\\
-1.0 & 0.5354 & 0.8865 & 15.130 & 42.060\\
-1.5 & 0.4981 & 0.9101 & 15.716 & 53.992\\
\end{tabular}
\end{table}

\begin{figure}
\caption[ ]{The $B$ dependence (in $-\chi Q\cdot
Q(1+B\vec{\tau}(1)\cdot \vec{\tau}(2))$) of the energies of selected
states in $^{10}Be$ in large space ($0p$ shell $+2 \hbar \omega$
excitations). These are: \\

{\bf (a)} Dashed curve: the $2_1^+$ state.\\
{\bf (b)} Solid curve: the $L=1~S=1$ triplet with $J=0^+,1^+,2^+$.\\
{\bf (c)} Dashed curve and circle: the most strongly excited $2^+$
state.\\
{\bf (d)} Dash-Dot curve: the $J=1^+~T=1$ scissors mode
($L=1~S=0$).\\
{\bf (e)} Dotted curve: the $J=1^+~T=2$ scissors mode.}
\end{figure}

\begin{figure}
\caption{The $B$ dependence of the energies of the
$2_1^+~T=0$ state (dashed curve), and the scissors mode state
($L=1~S=0~T=1$) (solid curve) in $^8Be$ (large space).}
\end{figure}

\begin{figure}
\caption[ ]{The `two-particle problem' $^6He$
($^6Be$). Excitation energies of selected states in a small space ($0p$
shell) calculation:\\

{\bf (a)} Solid curve: the $L=0~S=0~J=0^+$ state.\\
{\bf (b)} Dashed curve: the $L=2~S=0~J=2^+$ state.\\
{\bf (c)} Dot-dashed curve: the $L=1~S=1$ triplet $J=0^+,1^+,2^+$.}
\end{figure}

\begin{figure}
\caption[ ]{The $B$ dependence of the energies of selected
states in $^{22}Ne$ in a small space ($0d-1s$ shell) calculation:\\

{\bf (a)} Dashed curve: the $2_1^+$ state.\\
{\bf (b)} Dashed-circle curve: the $2^+$ state with the largest
$B(E2)$ from the ground state.\\
{\bf (c)} Solid curve: the $L=1~S=1$ triplet $J=0^+,1^+,2^+$.\\
{\bf (d)} Dash-Dot curve -the $J=1^+~T=1$ scissors mode
($L=1~S=0$).\\
{\bf (e)} Dotted curve: the $T=2$ branch of the scissors mode.}
\end{figure}

\end{document}